\documentclass[twocolumn,showpacs,preprintnumbers,amsmath,amssymb]{revtex4}

\usepackage[dvips]{graphicx}
\usepackage[tight]{subfigure}
\usepackage{subfigure}

\begin{document}

\author{T. Kroll$^1$}
\email{t.kroll@ifw-dresden.de}
\author{M. Knupfer$^{1}$}
\author{J. Geck$^{1}$}
\author{C. Hess$^{1}$}
\author{T. Schwieger$^{1}$}
\author{G. Krabbes$^{1}$}
\author{C. Sekar$^{1}$}
\author{D.R. Batchelor$^{2}$}
\author{H. Berger$^{3}$}
\author{B. B\"{u}chner$^{1}$}

\affiliation{$^{1}$ IFW Dresden, P.O. Box 270016, D-01171 Dresden,
Germany} \affiliation{$^{2}$ Universit\"at W\"urzburg, Am Hubland,
D-97074 W\"urzburg, Germany} \affiliation{$^{3}$ Institute of
Physics of Complex Matter, EPFL, CH-1015 Lausanne, Switzerland}

\title{X-ray absorption spectroscopy on layered cobaltates $\rm\bf Na_xCoO_2$}

\begin{abstract}
Measurements of polarization and temperature dependent soft x-ray
absorption have been performed on $\rm Na_xCoO_2$ single crystals
with x=0.4 and x=0.6. They show a deviation of the local trigonal
symmetry of the $\rm CoO_6$ octahedra, which is temperature
independent in a temperature range between 25 K and 372 K. This
deviation was found to be different for $\rm Co^{3+}$ and $\rm
Co^{4+}$ sites. With the help of a cluster calculation we are able
to interpret the Co $\rm L_{23}$--edge absorption spectrum and
find a doping dependent energy splitting between the $t_{2g}$ and
the $e_g$ levels (10Dq) in $\rm Na_xCoO_2$.
\end{abstract}

\maketitle

\section{Introduction}
The discovery of an unexpectedly large thermopower accompanied by
low resistivity and low thermal conductivity in $\rm Na_xCoO_2$
raised significant research interest in these materials
\cite{Terasaki_PRB97} and lead to a number of experimental and
theoretical investigations \cite{Ray_PRB99, Koshibae_PRB00,
Singh_PRB00, Motohashi_PRB03, Wang_Nature03}. This interest has
strongly been reinforced by the discovery of superconductivity in
the hydrated compound $\rm Na_{0.35}CoO_2\cdot 1.3 H_2O$ in 2003
\cite{Takada_Nature03}, and thus $\rm Na_xCoO_2$ experiences an
again increasing attention \cite{Lorenz_PRB03, Jorgensen_PRB03,
Baskaran_PRL03, Singh_PRB03, Chen_PRB04, Li_PRL04, Chainani_PRB04,
Koshibae_PRL03,Wu_PRL05, Bernhard_PRL04, Kubota_PRB04,
Mikami_JJAP03, Bayrakci_PRB04, Boothroyd_PRL04, Indergand_PRB05,
Mizokawa_PRB05, Foo_PRL04}. The similarity of the Na cobaltates to
the high temperature superconductors (HTSC) - both are transition
metal oxides and adopt a layered crystal structure with quasi two
dimensional $\rm (Cu,Co)O_2$ layers - is an important aspect of
the research activities. In contrast to the HTSC cuprates however,
the $\rm CoO_2$ layers consist of edge sharing $\rm CoO_6$
octahedra which are distorted in a way that the resulting symmetry
is trigonal. The trigonal coordination of the Co--sites results in
geometric frustration which favors unconventional electronic
ground states. The geometrically frustrated $\rm
CoO_2$--sublattice also exists in the non--hydrated parent
compound $\rm Na_xCoO_2$, which has been investigated in this
work. The intercalation of water into the parent compound is
expected to have little effect on the Fermi surface beside the
increase in two--dimensionality due to the effect of expansion
\cite{Rosner_BJP03,Johannes_PRB04}.

\par

Upon lowering the symmetry from cubic to trigonal, the $t_{2g}$
states split into states with $\rm e_g^\pi$ and $\rm a_{1g}$
symmetry, which can be represented as
\begin{equation}
\rm e_{g\pm}^\pi=\mp\frac{1}{\sqrt{3}}[|xy\rangle+\exp^{\pm
i2\pi/3}|yz\rangle+\exp^{\pm i4\pi/3}|xz\rangle]
\end{equation}
and
\begin{equation}
\rm a_{1g}=\frac{1}{\sqrt{3}}[|xy\rangle+|yz\rangle+|xz\rangle].
\label{a1g state}
\end{equation}
The cubic $e_g$ states remain degenerate and will be named
$e_g^{\sigma}$ in the following in order to avoid confusion. As
has been predicted by calculations \cite{Koshibae_PRL03} and shown
experimentally in Ref. \onlinecite{Wu_PRL05}, the $\rm
e_g^\pi$-states are lower in energy and are therefore filled,
while the $\rm a_{1g}$-states are partially filled as a function
of x. Therefore, the $\rm a_{1g}$--states govern the relevant low
energy excitations.
\begin{figure}[!]
\begin{center}
\includegraphics[width=0.8\columnwidth, angle=0, clip]{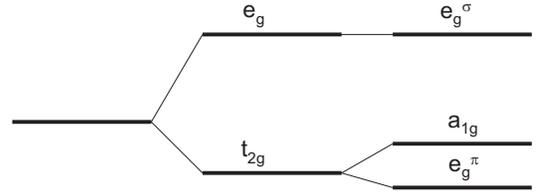}
\end{center}
\caption{\small Illustration of the splitting of the energy levels
for cubic and trigonal symmetry. Here due to the trigonal
distortion the $e_g^{\pi}$ states have been assumed to be lowest
in energy.} \label{aufspaltung}
\end{figure}

\par

From magnetization, specific heat, and conductivity measurements
for various doping levels, three phase transitions as a function
of temperature have been observed: a low temperature bulk
antiferromagnetic transition at $\rm T_N\approx 20$ K for $\rm
0.75\le x \le 0.9$ \cite{Motohashi_PRB03, Bayrakci_PRB04,
Mikami_JJAP03} and two high temperature transitions at about 280 K
in a $\rm Na_{0.82}CoO_2$ sample \cite{Bernhard_PRL04} and
323--340 K in samples with $\rm x\approx 0.75$ \cite{Sales_PRB04,
Huang_PRB04a}. The origin of the 280 K transition has been
discussed in terms of charge ordering and the formation of
magnetopolarons due to a charge induced Co 3$d$ spin state
transition from a low spin state to an intermediate spin state
\cite{Bernhard_PRL04}. Such a spin state transition occurs for
instance in the related compounds $\rm La_{1-y}Sr_yCoO_3$
\cite{Toulemonde_JSSC01, Zobel_PRB02, Yamaguchi_PRB96,
Loshkareva_PRB03}. In some samples, a third transition at around
340 K is observed, and it was suggested that this is due to a
structural transition involving Na ordering \cite{Sales_PRB04,
Huang_PRB04a, Geck_CM05}, which is supported by Huang {\it et al.}
who found a structural transition concomitant with a shift of a
large fraction of the Na ions from a high--symmetry position to a
lower--symmetry position \cite{Huang_PRB04a}.

\par

In this article, we present measurements of the near edge x--ray
absorption fine structure (NEXAFS) of $\rm Na_xCoO_2$ ($\rm x=0.4$
and $\rm x=0.6$) which have been carried out in order to
investigate the electronic properties of this interesting class of
materials.

\section{Experimental}

\label{section experimental}

The used single crystals were grown by two different methods.
First, single crystals of $\rm Na_{0.75}CoO_2$ were grown using
the travelling solvent floating--zone method, crystals with a
lower Na concentration were produced using de--intercalation with
Bromium from the highly doped samples. The initial
characterization of the samples has been carried out using energy
dispersive x-ray analysis (EDX), x--ray diffraction and chemical
analysis. Details of the crystal growth, de--intercalation, and
characterization of the resulting samples will be presented
elsewhere \cite{Shekar_xy}. Second, high--quality single crystals
were grown by the sodium chloride flux method as thoroughly
described in Ref. \onlinecite{Iliev_PC04}. Both methods lead to
the same spectra.

\par

The NEXAFS measurements of the absorption coefficient were
performed at the UE52-PGM beamline at the synchrotron facility
BESSY II, Berlin, analyzing the drain current. The energy
resolution was set to 0.09 eV and 0.16 eV for photon energies of
530 eV and 780 eV, respectively. We performed measurements on
different non--hydrated single crystals with a sodium content of
$\rm x=0.4$ and $\rm x=0.6$ at various temperatures and different
polarizations of the incident synchrotron light at the oxygen $K$-
and cobalt $L_{2,3}$--edge. The two doping levels that have been
investigated belong to two different and interesting regions in
the phase diagram \cite{Foo_PRL04}, x=0.4 lies in the region of a
paramagnetic metal whereas and x=0.6 is a Curie-Weiss metal. All
the crystals were of the same size of about 3x3 $\rm mm^2$ area
and 1 mm thickness. Crystals were freshly cleaved in--situ under
ultra--high vacuum conditions (about $\rm 2\cdot 10^{-10}$ mbar)
at 25 K, which resulted in shiny sample surfaces.

\par

Special attention has been paid to the reproducibility of the
experimental data and the effects of surface contaminations. The
freshly cleaved surfaces turned out to be very sensitive to
adsorbates. We observed irreversible changes in the O $K$--spectra
when the sample surfaces were exposed to pressures above $\rm
2\cdot 10^{-9}$ mbar which we attribute to adsorbed, oxygen
containing molecules at the surface. No such changes have been
found for the Co absorption edges. A comparison to the temperature
dependent behavior of the Co $L_{2,3}$--edge proves that the
origin of this irreversible effect is due to surface contamination
and not due to a change or a transition of the whole sample, since
the Co $L_{2,3}$--edge remains unchanged with temperature.

\par

According to the dipole selection rules the O {\it K} and Co
$L_{2,3}$ excitations as probed by these experiments, correspond
to core electron transitions into unoccupied oxygen 2{\it p} and
cobalt 3{\it d} electronic states. Upon variation of the incident
light polarization, different O 2{\it p} and Co 3{\it d} orbitals
can be probed \cite{fink1994}. For polarization dependent
measurements, the samples were oriented such that the direction of
incident photons and the sample surface normal (i.e. the
$c$--axis) enclose an angle of $\rm \alpha =70^\circ$. The used
undulator allows a rotation of the beam polarization by $\rm
90^\circ$ using the vertical and horizontal mode. This procedure
avoids experimental artifacts related to the differences in the
optical path and the probed area. All NEXAFS results referred to
$E$ parallel to the $c$--axis ($\rm E||c$) are corrected using the
formula $I_{||c}=\frac{1}{\sin^2 (\alpha)}(I-I_{\perp c}\cos^2
\alpha)$ where $I_{\perp c}$ and $I$ are measured NEXAFS
intensities with $E\perp c$ and $E$ in the plane defined by the
$c$--axis and the incident photon beam, respectively. In order to
compare the data, the NEXAFS spectra are normalized at higher
energies where the absorption is doping independent and isotropic
and the spectra for different measurements and settings should
show the same intensities. The spectra are normalized at 600 eV
for the O {\it K}--edge and at 810 eV for the Co $L_{2,3}$--edge.

\section{Results and Discussion}

Using NEXAFS, the unoccupied energy levels close to the Fermi
level can be studied by excitations from core electrons into
unoccupied states. We have measured excitations from  O 1{\it s}
core levels into unoccupied O 2{\it p} states that are hybridized
with states of primary Co and Na character, as well as excitations
from Co 2{\it p} into Co 3{\it d} states. If the influence of the
core hole is neglected, as is reasonable for the O 1{\it s} core
hole excitations, a direct interpretation of the NEXAFS results in
terms of the partial unoccupied density of states is possible,
analogous to a one electron addition process
\cite{deGroot_PRB89,deGroot_CCR04}. If the core hole cannot be
neglected, as is the case for Co 3{\it d} excitations, the
interpretation of the data requires consideration of multiplet
splitting, hybridization and crystal field effects.

\subsection{Co {\it L}-edge}

\label{Co part}

\begin{figure}
\begin{center}
\includegraphics[bb=10 10 359 267,width=0.9\columnwidth,angle=0,clip]{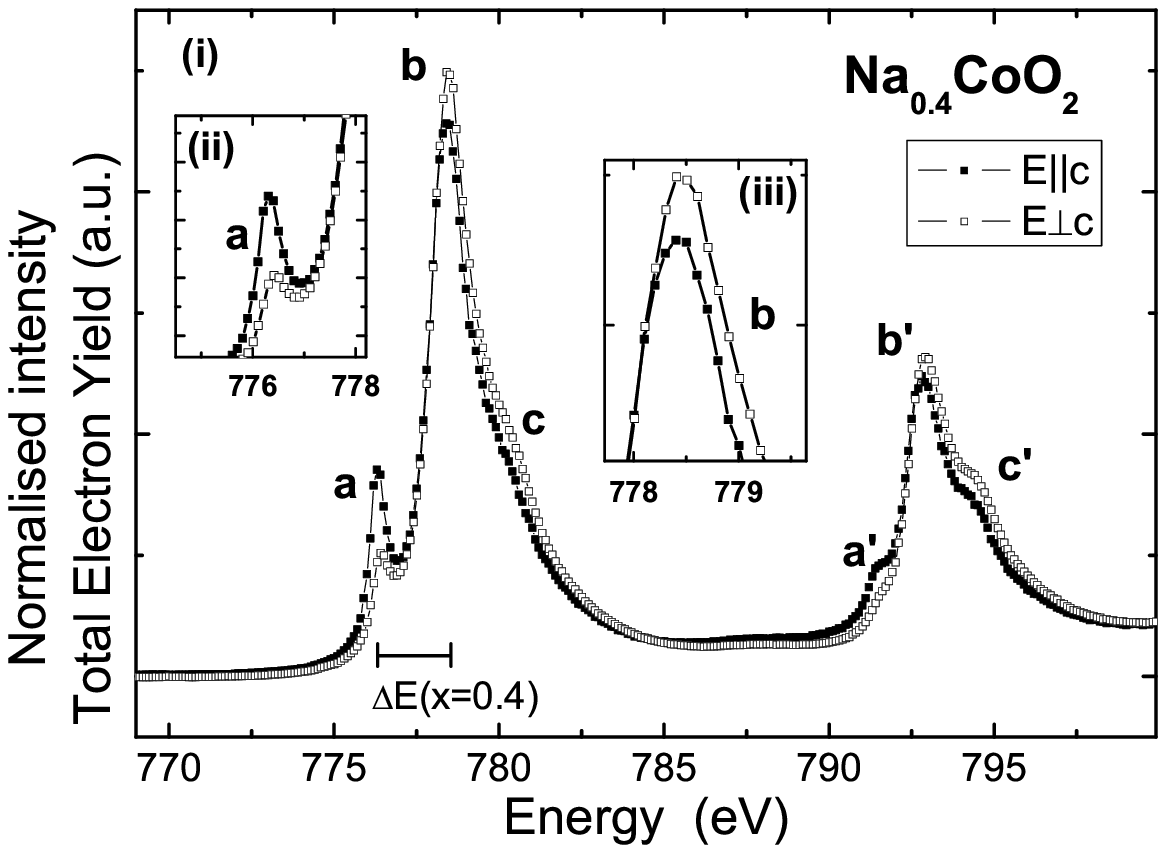}
\includegraphics[bb=10 10 359 272,width=0.9\columnwidth,angle=0,clip]{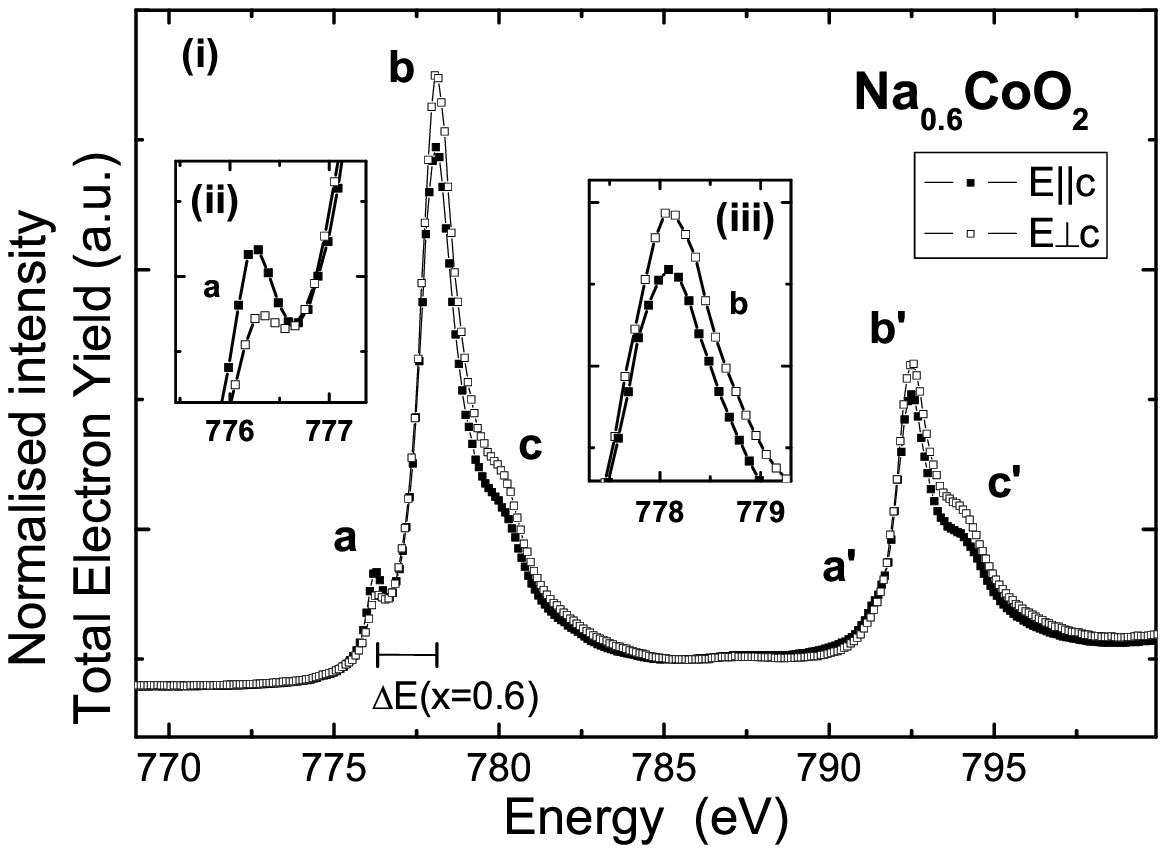}
\includegraphics[bb=10 10 273 225,width=0.9\columnwidth,angle=0, clip]{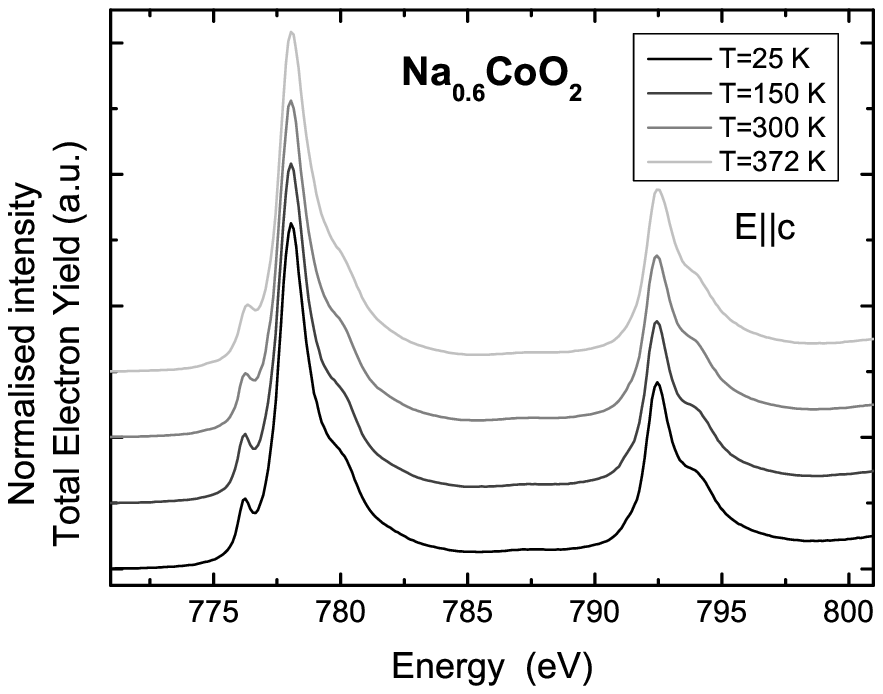}
\end{center}
\caption{\small Top and middle figure: Polarization dependence
NEXAFS spectra of the Co $\rm L_{2,3}$--edge for $\rm Na_xCoO_2$
with x=0.4 (top) and x=0.6 (middle) at 25 K. $\rm \Delta E_x$
represents the difference in energy between the first and the main
peak. The insets show an enlargement of the region around the
first peak (ii) and around the top of the main peak (iii). Bottom:
No significant temperature dependence is observable between 25 and
372 K for x=0.6. All intensities are normalized at 810 eV where no
stoichiometric, polarization, or temperature dependence is
observable.} \label{Co2p(T)}
\end{figure}

In Fig. \ref{Co2p(T)} experimental spectra of the Co
$L_{2,3}$--edges are shown, which display three main features at
each edge: one strong central peak (peak b and b') with a shoulder
towards higher energies (peak c and c') and a peak/shoulder
towards lower energies (peak a and a') (Fig. \ref{Co2p(T)}). This
result can easily be compared to the similar compound $\rm
LiCoO_2$ which nominally contains only $\rm Co^{3+}$ ions with S=0
\cite{Johnston_JPCS58}. In the NEXAFS spectra of the Co $\rm
L_{2,3}$--edge of $\rm LiCoO_2$ one finds only one main peak
\cite{vanElp_PRB91} different from the spectrum observed for the
mixed valence system $\rm Na_xCoO_2$. Especially the low energy
feature (peak a and a') is absent in $\rm LiCoO_2$, consequently
we assign this peak to be caused by excitations into unoccupied
3{\it d} states of nominal $\rm Co^{4+}$ ions which are missing in
$\rm LiCoO_2$. This interpretation is furthermore supported by the
doping dependence of the {\it L}--edge as the low energy peak
appears stronger for lower sodium doping (i.e. higher $\rm
Co^{4+}$ concentration) and weaker for higher sodium doping (i.e.
lower $\rm Co^{4+}$ concentration), as well as by calculations as
described below.

\par

The local electronic structure around a Co atom in $\rm Na_xCoO_2$
has been modelled in a cluster calculation using many--body wave
functions. Within this approach, a $\rm CoO_6$ cluster containing
the Co 3$d$ and the O 2$p$ valence electrons has been solved
exactly including all interactions between 3$d$ electrons in $O_h$
symmetry with a ground state for $\rm Co^{4+}$ as given in
equation \ref{a1g state} \cite{Kroll_06}. For simplicity we will
use in the following the expression $\rm Co^{4+}$, referring to a
(distorted) $\rm CoO_6$ octahedra with a formal $\rm Co^{4+}$
central ion containing five holes and, analogously, $\rm Co^{3+}$
for an octahedra containing four holes with a formal $\rm Co^{3+}$
central ion. It has been found from the calculations that the
first peak at lower energies in the NEXAFS Co $L$--edge (peak a
and a' in Fig. \ref{Co2p(T)}) originates from excitations into
$\rm Co^{4+}$ final states with an $\rm A_{1g}$ symmetry, while
the main peak (b and b') and the shoulder (c and c') are due to
excitations into final states of $\rm Co^{3+}$ with $\rm E_g$
symmetry and $\rm Co^{4+}$ with $\rm T_{1g}$ and $\rm T_{2g}$
symmetry, respectively. Note that in order to avoid confusion we
labeled the final states by using capital letters (e.g. $A_{1g}$)
while the ground is labeled with lower case letter (e.g.
$a_{1g}$). The ground state of the system has been found to be
strongly covalent with a moderate positive charge transfer energy
$\rm \Delta_{CT} = E(d^{n+1}L)-E(d^{n})$ for $\rm Co^{3+}$ and a
negative $\rm \Delta_{CT}$ for $\rm Co^{4+}$ \cite{Kroll_06}.

\par

In the experimental spectrum of the Co {\it L}--edge one
additionally observes that the energy difference  between the
first peak ($A_{1g}$) and the largest peak ($E_g$) in Fig.
\ref{Co2p(T)} differs between the two different stoichiometries
x=0.4 and x=0.6 being $\rm \Delta E_{x=0.4}=-2.0$ eV and $\rm
\Delta E_{x=0.6}=-1.8$ eV. Theoretically, this behavior can be
best explained by a change of the energy splitting between the $
t_{2g}$ states and the $e_g$ states (10Dq). An explanation for a
lower 10Dq for higher sodium intercalation can be found from
neutron powder diffraction by Huang {\it et al}. An increasing
Co--O distance with increasing sodium content is found
\cite{Huang_PRB04a} which guides a lower influence of the crystal
field and charge transfer, i.e. a lower 10Dq.

\par

The $\rm a_{1g}$ orbital of the ground state in trigonal symmetry
points along the (1 1 1) direction of the distorted $\rm CoO_6$
octahedra, i.e. parallel to the crystal $c$--axis, while the the
two $e_g^{\pi}$ orbitals point perpendicular to it. From
polarization dependent measurements with the {\bf E} vector of
photons parallel and $\rm 70^{\circ}$ to the crystal $c$ axis one
observes the $\rm a_{1g}$ peak of $\rm Co^{4+}$ to be stronger for
$\rm {\bf E}||c$ as compared to $\rm {\bf E}\perp c$. This
behavior is expected for a local trigonal distortion, where the
$\rm t_{2g}$ ground states split into states with $\rm a_{1g}$ and
$\rm e_{g}^{\pi}$ symmetry, whereas the $\rm e_{g}^{\sigma}$
states remain untouched (c.f. Fig. \ref{aufspaltung}) and
therefore should not show a strong polarization dependence.
Different from that, the intensity of the $\rm Co^{3+}$ central
peak is significantly larger for $\rm {\bf E}\perp c$ compared to
$\rm {\bf E}||c$ (Fig. \ref{Co2p(T)}). This effect points to an
additional distortion that splits the $\rm e_g^{\sigma}$ levels.
Such a splitting might be caused by two mechanisms. From spectral
ellipsometry on a $\rm Na_{0.82}CoO_2$ sample, Bernhard {\it et
al.} find a transition at 280 K which they explain by the
formation of magnetopolarons using the idea of a spin--state
transition \cite{Bernhard_PRL04} similar to the related compound
$\rm La_{1-y}Sr_yCoO_3$ \cite{Toulemonde_JSSC01, Zobel_PRB02,
Yamaguchi_PRB96, Loshkareva_PRB03}. In $\rm Na_xCoO_2$, this
mechanism would be driven by a displacement of the neighboring
oxygen ligands towards the central $\rm Co^{4+}$ ion which may
favor an intermediate--spin (IS) state with S=1 over a low--spin
(LS) state with S=0 of the $\rm Co^{3+}$ ions
\cite{Bernhard_PRL04}. This displacement would result in a
splitting of the $\rm e_g^{\sigma}$ levels. Another possibility
for an additional distortion could arise from the effect of sodium
ordering at special doping levels \cite{Huang_PRB04a,
Zandbergen_PRB04, Zhang_PRB05, Geck_CM05}, it is assumed that this
results in orthorombic symmetry so that a perturbation of the
trigonal distortion of the octahedra could occur.

Both possible contributions are related to a corresponding
ordering temperature, the formation of magnetopolarons is observed
at 280 K, whereas the sodium ordering appears at temperatures
below 350 K for x=0.75. However, no significant temperature
dependence for $\rm Na_{0.6}CoO_2$ has been found between 25 K and
370 K in the present NEXAFS studies. This implies that the
spin--state of Co as well as the electronic structure in this
energy range as seen by NEXAFS is not affected by temperature
neither by a spin state transition nor by a structural transition
involving Na ordering (Fig. \ref{Co2p(T)} bottom). A temperature
independent distortion would be expected if the non--trigonal
distortion is purely structural.

\subsection{O {\it K}-edge}

\begin{figure}[!]
\begin{center}
\includegraphics[width=1.0\columnwidth, angle=0, clip]{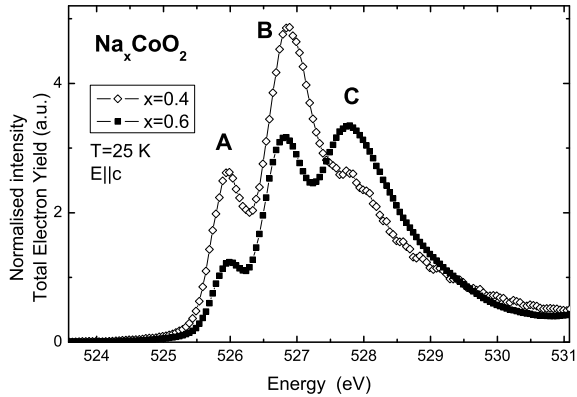}
\end{center}
\caption{\small Stoichiometric dependence of $\rm Na_xCoO_2$ at
the Oxygen $K$-edge. A, B, and C represent the positions of the
three main peaks. The intensities are normalized at 600 eV.}
\label{O1s(x)}
\end{figure}
\begin{figure}[!]
\begin{center}
\includegraphics[width=1.0\columnwidth, angle=0, clip]{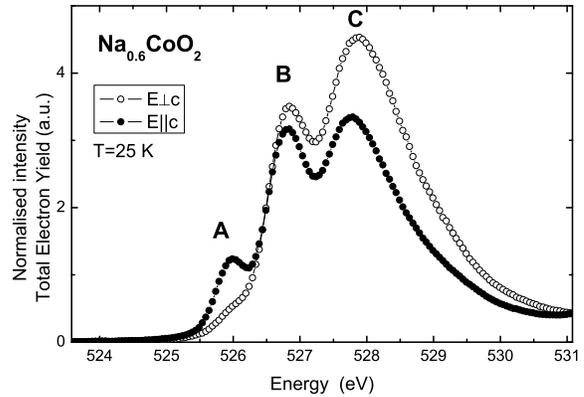}
\includegraphics[width=1.0\columnwidth, angle=0, clip]{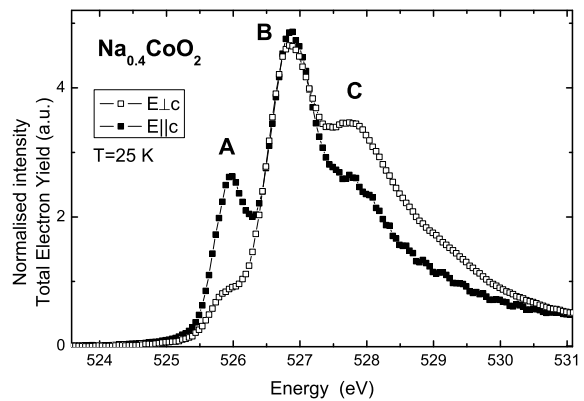}
\end{center}
\caption{\small NEXAFS polarization dependence of the oxygen
$K$-edge of $\rm Na_xCoO_2$ for x=0.6 (top) and x=0.4 (bottom).
Polarization $E||c$ is indicated by filled symbols, $E||c$ is
indicated by open symbols. The intensities are normalized at 600
eV where the spectrum is isotropic.} \label{O1s(p)}
\end{figure}

In Figure \ref{O1s(x)}, we present the results for the O $K$
absorption edge of $\rm Na_{0.4}CoO_2$ and $\rm Na_{0.6}CoO_2$
measured with light polarized parallel to the crystallographic $c$
axis. For both stoichiometries we observed three pronounced
features A, B and C above the absorption threshold, showing a
significant doping dependence. Features A and B increase for
smaller x, while feature C decreases, i.e. features A and B
increase with an increasing hole doping. In the related compound
$\rm LiCoO_2$ the situation is very similar to $\rm Na_1CoO_2$,
since both systems are assumed to have a low--spin state meaning
that all six $t_{2g}$ states are occupied by electrons while the
four $e_g$ states are empty. The resulting absorption spectrum
shows only one peak due to O 2$p$ orbitals hybridized with the Co
3$d$ orbitals with $\rm e_g$ symmetry \cite{vanElp_PRB91}. In $\rm
Na_xCoO_2$, excitations into unoccupied $\rm Co^{3+}$ states
should therefore only be responsible for a single peak in the
whole O {\it K}-absorption edge spectra of $\rm Na_xCoO_2$. We
therefore attribute the first two features A and B in Fig.
\ref{O1s(x)} to doping induced states related to the formation of
$\rm Co^{4+}$ sites and feature C to the formation of of $\rm
Co^{3+}$ sites, surrounded by oxygen octahedra. It has been shown
previously that the pre--edge peaks in the O $K$ NEXAFS spectra of
the late transition--metal (TM) oxides are shifted by about 1 eV
to lower energies when the TM valence increases by 1
\cite{Hu_ChemPhys98}; therefore, spectral features from two
different valence states can be well resolved. The energetic
downshift can be explained by a decrease of the energy between the
partially filled Co 3{\it d} states; the valence electrons will be
screened by 1 eV more with every added valence electron, resulting
in a situation that for late transition metals a higher valency
corresponds to a lower excitation energy. Upon hole addition
(decreasing x), the increase of features A and B, and the decrease
of feature C indicate that both are related to the doping process
and it is natural to ascribe them to excitations into unoccupied O
2$p$ states hybridized with $\rm Co^{3+}$ $\rm e_g^\sigma$ states
(feature C) and, about 1 eV lower in energy, to excitations into
unoccupied O 2$p$ states hybridized with $\rm Co^{4+}$ $\rm
e_g^\sigma$ states (feature B) and those hybridized with $\rm
Co^{4+}$ $\rm a_{1g}$ states (feature A). Additionally, these
results show that the holes in $\rm Na_{x}CoO_2$ have a
significant oxygen character, which is in good agreement with
other cobalt based compounds \cite{vanElp_PRB91, Hu_PRL04,
Yamaguchi_PRB96}.

\par

Next we turn to the polarization dependence of the O $K$
absorption edges as shown in Fig. \ref{O1s(p)}. From polarization
dependent absorption measurements, information about the
orientation of the corresponding orbitals are obtained. Our data
signal that the doping induced absorption feature A is strong for
$\rm E||c$, and substantially weaker for $\rm E\perp c$.
Consequently, from the orientation of the $a_{1g}$--orbital and
the attribution of the three peaks as described above, the holes
doped into the $\rm CoO_2$ layers of $\rm Na_xCoO_2$ have a
predominant $\rm a_{1g}$ character similar to the result found by
Wu {\it et al.} using x--ray absorption spectroscopy
\cite{Wu_PRL05}. Band structure calculations find, that although
the $a_{1g}$ and $e_g^{\pi}$ states overlap, and mix to some
extend, the centers of these states are energetically separated
with those closer to $\rm E_F$ having dominant $a_{1g}$ character
\cite{Singh_PRB00}, in good agreement with our results. In
addition, while feature B is not or only slightly polarization
dependent, the intensity of peak C ($\rm Co^{3+}$) is
significantly stronger for ${\bf E}\perp c$ than for ${\bf E}||
c$. The same result has already been observed at the Co $L$--edge,
but because of the large central peak no quantitative analysis of
the polarization dependence of the $\rm Co^{4+}$ shoulder could be
made. At the oxygen $K$--edge these two peaks are well separated
and a difference in the response due to different polarizations is
observable. From this we conclude that the trigonal symmetry is
better realised in octahedra containing a $\rm Co^{4+}$ central
ion than in octahedra containing a $\rm Co^{3+}$ central ion. Such
a situation has already been suggested by Bernhard and coworkers
as an origin of magnetopolarons \cite{Bernhard_PRL04}. As already
mentioned in section \ref{Co part}, the underlying idea for such
an effect is a lowering of the local symmetry around the $\rm
Co^{3+}$. Thus, the $t_{2g}$ triplet and the $e_g$ doublet split
and become polarization dependent. The effect may be stronger at
temperatures below 20 K, which is also the critical temperature
for a magnetic transition to a bulk antiferromagnetic ordered
state \cite{Bayrakci_PRB04, Foo_PRL04}, at temperatures lower than
20 K the mobility of the magnetopolarons is assumed to be lower
due to an increased self--trapping energy in the antiferromagnetic
state \cite{Bernhard_PRL04}.

\begin{figure}[t]
\begin{center}
\includegraphics[width=1.0\columnwidth, angle=0, clip]{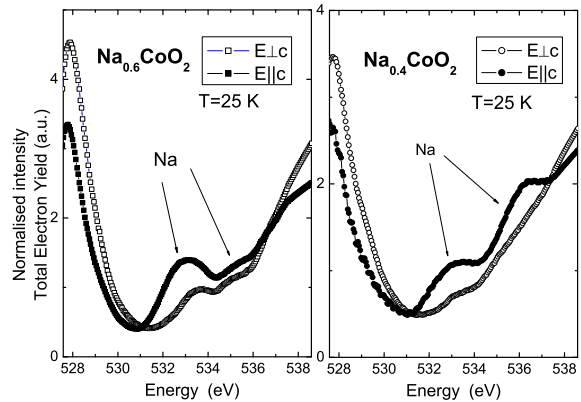}
\end{center}
\caption{\small Stoichiometric and polarization dependence of the
NEXAFS spectra of $\rm Na_xCoO_2$ at the oxygen $K$--edge. The
energy range right above the oxygen threshold is shown where the
the hybridization between O and Na can be monitored. All curves
are normalized at 600 eV.} \label{Na}
\end{figure}

Somewhat higher in energy at $\rm E \approx 535$ eV, one finds the
excitations into unoccupied O levels which are hybridized with Na
orbitals \cite{Wu_PRL05}. As expected, the resulting peaks
increase in intensity with increasing sodium intercalation, but in
addition they are strongly polarization dependent. As is shown in
Fig. \ref{Na} the intensity for $ \vec{E}||c$ is stronger than for
$ \vec{E}\perp c$. This leads to a finite Na--O hybridization
along $c$, a consequence of this hybrid could be a 3 dimensional
magnetism as has been proposed by Johannes {\it et al.}
\cite{Johannes_PRB05}. From our data it becomes clear, that the
inter--planar binding is more likely to have a covalent rather
than an ionic character, so that a 3D magnetism is reasonable.

\section{Conclusion and summary}

The study of single crystals with a stoichiometry $\rm
Na_{0.4}CoO_2$ and $\rm Na_{0.6}CoO_2$ reveal polarization
dependencies that cannot be explained by a simple trigonal
distortion of the $\rm CoO_6$ octahedra. Taking the results of
both, the Co $L_{2,3}$--edge and the O $K$--edge into account, it
follows that an additional distortion is present which is stronger
for octahedra with a formal $\rm Co^{3+}$ central ion than for
octahedra with a formal $\rm Co^{4+}$ central ion. A possible
explanation for such a phenomena has been given by Bernhard {\it
et al.} who find magnetopolarons in $\rm Na_{0.82}CoO_2$ due to a
lowering of the local symmetry \cite{Bernhard_PRL04}. Furthermore,
we find doping dependent relative peak positions at the Co
$L_{2,3}$--edge which can be explained by a doping dependent
splitting of the $t_{2g}$--$e_g$ levels (10Dq) as has been found
from cluster calculations and can be explained by a doping
dependent Co--O bond distance \cite{Huang_PRB04a}. At the O
$K$--edge we find excitations into unoccupied O levels which are
hybridized with Na orbitals to be strongly polarization dependent,
which emphasizes the covalent character of the inter--planar
binding of the $\rm CoO_2$ and Na planes, rather than an ionic
character.

\section*{Acknowledgment}
We are grateful to R. H\"ubel, S. Leger and R. Sch\"onfelder for
technical assistance and J. Acker for the chemical analysis. This
investigation was supported by the DFG project KL 1824/2 and KR
1241/3 4, and the Deutscher Akademischer Austauschdienst (DAAD).


\begin{thebibliography}{99}
\bibitem{Terasaki_PRB97} I. Terasaki, Y. Sasago, and K. Uchinokura, Phys. Rev. B {\bf 56} R12685 (1997)
\bibitem{Ray_PRB99} R. Ray, A. Ghoshray, and K. Ghoshray, and S. Nakamura, Phys. Rev. B {\bf 59} 9454 (1999)
\bibitem{Koshibae_PRB00} W. Koshibae, K. Tsutsui, and S. Maekawa, Phys. Rev. B {\bf 62} 6869 (2000)
\bibitem{Singh_PRB00} D.J. Singh, Phys. Rev. B {\bf 61} 13397 (2000)
\bibitem{Motohashi_PRB03} T. Motohashi, R. Ueda, E. Naujalis, T. Tojo, I. Terasaki, T. Atake, M. Karppinen,
and H. Yamauchi, Phys. Rev. B {\bf 67} 064406 (2003)
\bibitem{Wang_Nature03} Y. Wang, N.S. Rogado†, R.J. Cava, and N. P. Ong, Nature {\bf 423} 425 (2003)
\bibitem{Takada_Nature03} K. Takada, H. Sakurai, E. Takayama-Muromachi,
F. Izumi, R.A. Dilanian, and T. Sasaki, Nature {\bf 422} 53 (2003)
\bibitem{Lorenz_PRB03} B. Lorenz, J. Cmaidalka, R.L. Meng, and C.W. Chu, Phys. Rev. B {\bf 68} 132504 (2003)
\bibitem{Jorgensen_PRB03} J.D. Jorgensen, M. Avdeev, D.G. Hinks, J.C. Burley, and S. Short,
Phys. Rev. B {\bf 68} 214517 (2003)
\bibitem{Baskaran_PRL03} G. Baskaran, Phys. Rev. Lett. {\bf 91} 097003 (2003)
\bibitem{Singh_PRB03} D.J. Singh, Phys. Rev. B {\bf 68} 20503 (2003)
\bibitem{Chen_PRB04} D.P. Chen, H.C. Chen, A. Maljuk, A. Kulakov, H. Zhang, P. Lemmens, and C. T. Lin,
Phys. Rev. B {\bf 70} 024506 (2004)
\bibitem{Li_PRL04} S.Y. Li, L. Taillefer, D.G. Hawthorn, M.A. Tanatar, J. Paglione, M. Sutherland,
R.W. Hill, C.H.Wang, and X.H. Chen, Phys. Rev. Lett. {\bf 93}
056401 (2004)
\bibitem{Chainani_PRB04} A. Chainani, T. Yokoya, Y. Takata, K. Tamasaku, M. Taguchi, T. Shimojima,
N. Kamakura, K. Horiba, S. Tsuda, S. Shin, D. Miwa, Y. Nishino, T.
Ishikawa, M. Yabashi, K. Kobayashi, H. Namatame, M. Taniguchi, K.
Takada, T. Sasaki, H. Sakurai, and E. Takayama-Muromachi, Phys.
Rev. B {\bf 69} 180508 (2004)
\bibitem{Koshibae_PRL03} W. Koshibae and S. Maekawa, Phys. Rev. Lett. {\bf 91} 257003 (2003)
\bibitem{Wu_PRL05} W.B. Wu, D. J. Huang, J. Okamoto, A. Tanaka, H.J. Lin, F.C. Chou, A. Fujimori,
and C.T. Chen, Phys. Rev. Lett.  {\bf 94} 146402 (2005)
\bibitem{Bernhard_PRL04} C. Bernhard, A.V. Boris, N.N. Kovaleva, G. Khaliullin, A.V. Pimenov, Li Yu,
D.P. Chen, C.T. Lin, and B. Keimer, Phys. Rev. Lett. {\bf 93}
167003 (2004)
\bibitem{Kubota_PRB04} M. Kubota, K. Takada, T. Sasaki, H. Kumigashira,
J. Okabayashi, M. Oshima, M. Suzuki, N. Kawamura, M. Takagaki, and
K. Ono, Phys. Rev. B {\bf 70} 12508 (2004)
\bibitem{Mikami_JJAP03} M. Mikami, M. Yoshimura, Y. Mori, Takatomo Sasaki,
R. Funahashi, and Masahiro Shikano, Jpn. J. Appl. Phys. {\bf 42}
7383–7386 (2003)
\bibitem{Bayrakci_PRB04} S.P. Bayrakci, C. Bernhard, D.P. Chen, B. Keimer, R.K. Kremer, P. Lemmens,
C.T. Lin, C. Niedermayer, and J. Strempfer, Phys. Rev. B {\bf 69}
100410 (2004)
\bibitem{Boothroyd_PRL04} A.T. Boothroyd, R. Coldea, D.A. Tennant, D. Prabhakaran, L.M. Helme,
and C.D. Frost, Phys. Rev. Lett. {\bf 92} 197201 (2004)
\bibitem{Indergand_PRB05} M. Indergand, Y. Yamashita, H. Kusunose, and M. Sigrist,
Phys. Rev. B {\bf 71} 214414 (2005)
\bibitem{Mizokawa_PRB05} T. Mizokawa, L.H. Tjeng, H.-J. Lin, C.T. Chen, R. Kitawaki, I. Terasaki,
S. Lambert, and C. Michel, Phys. Rev. B {\bf 71} 193107 (2005)
\bibitem{Foo_PRL04} M.L. Foo, Y. Wang, S. Watauchi, H.W. Zandbergen,
T. He, R.J. Cava, and N.P. Ong, Phys. Rev. Lett. {\bf 92} 247001
\bibitem{Johannes_PRB04} M.D. Johannes and D.J. Singh, Phys. Rev. B {\bf 70} 014507 (2004)
\bibitem{Rosner_BJP03} H. Rosner, S.-L. Drechsler, G. Fuchs, A. Handstein,
A. W{\"a}lte, K.-H. M{\"u}ller, Braz. Jour. Phys. {\bf 33} 718
(2003)
\bibitem{Sales_PRB04} B.C. Sales R. Jin, K.A. Affholter, P. Khalifah, G.M. Veith, and D. Mandrus,
Phys. Rev. B {\bf 70} 174419 (2004)
\bibitem{Huang_PRB04a} Q. Huang, B. Khaykovich, F.C. Chou, J.H. Cho, J.W. Lynn, and Y. S. Lee,
Phys. Rev. B {\bf 70} 134115 (2004)
\bibitem{Geck_CM05} J. Geck, M. v. Zimmermann, H. Berger, S.V. Borisenko, H. Eschrick, K. Koepernik,
M. Knupfer, and B. B{\"u}chner, cond-mat/0511554 (2005)
\bibitem{Toulemonde_JSSC01} O. Toulemonde, N. N.Guyen, F. Studer, and A. Traverse, Jour. of Solid State
Chem. {\bf 158} 208-217 (2001)
\bibitem{Zobel_PRB02} C. Zobel, M. Kriener, D. Bruns, J. Baier, M. Gr{\"u}ninger, T. Lorenz,
P. Reutler, and A. Revcolevschi, Phys. Rev. B {\bf 66} 020402
(2002)
\bibitem{Yamaguchi_PRB96} S. Yamaguchi, Y. Okimoto, H. Taniguchi, and Y. Tokura, Phys. Rev. B {\bf 53} 2926
(1996); {\bf 55} 8666 (1997)
\bibitem{Loshkareva_PRB03} N.N. Loshkareva, E.A. Ganshina, B.I. Belevtsev, Y.P. Sukhorukov,
E.V. Mostovshchikova, A.N. Vinogradov, V.B. Krasovitsky, and I. N.
Chukanova, Phys. Rev. B {\bf 68} 024413 (2003)
\bibitem{Shekar_xy} C. Sekar, G. Krabbes {\it et al.} to be published
\bibitem{Iliev_PC04} M.N. Iliev, A.P. Litvinchuk, R.L. Meng, Y.Y. Sun, J. Cmaidalka,
and C.W. Chu, Physica C {\bf 402} 239 (2004)
\bibitem{fink1994}J. Fink, N. N\"ucker, E. Pellegrin, H. Romberg,
M. Alexander, and M. Knupfer, J. Electron Spectosc. Relat. Phenom.
{\bf 66} 395 (1994)
\bibitem{deGroot_PRB89} F.M.F. de Groot {\it et al.}, Phys. Rev. B {\bf 40} 5715 (1989)
\bibitem{deGroot_CCR04} F.M.F. de Groot, Coor. Chem. Rev., {\bf 249} 31 (2005)
\bibitem{Johnston_JPCS58} W.D. Johnston, R.R. Heikes and D.
Sestrich, J. Phys. Chem. Solids, {\bf 7} 1 (1958)
\bibitem{vanElp_PRB91} J. van Elp, J.L. Wieland, H. Eskes, P. Kuiper, G.A. Sawatzky, F.M.F. de Groot, and
T.S. Turner, Phys. Rev. B {\bf 44} 6090 (1991)
\bibitem{Kroll_06} T. Kroll, A.A. Aligia, and G.A. Sawatzky, in preparation
\bibitem{Zandbergen_PRB04} H.W. Zandbergen, M. Foo, Q. Xu, V. Kumar, and R.J. Cava,
Phys. Rev. B {\bf 70} 024101 (2004)
\bibitem{Zhang_PRB05} P. Zhang, R.B. Capaz, M.L. Cohen, and S.G. Louie, Phys. Rev. B {\bf 71} 153102 (2005)
\bibitem{Hu_ChemPhys98} Z. Hu, G. Kaindl, S.A. Warda, D. Reinen, F.M.F. de Groot, and B.G. Muller,
Chemical Physics {\bf 232}
63-74 (1998)
\bibitem{Hu_PRL04} Z. Hu, Hua Wu, M.W. Haverkort, H.H. Hsieh, H.J. Lin, T. Lorenz, J. Baier, A. Reichl,
I. Bonn, C. Felser, A. Tanaka, C.T. Chen, and L.H. Tjeng, Phys.
Rev. Lett. {\bf 92} 207402 (2004)
\bibitem{Johannes_PRB05} M.D. Johannes, I.I. Mazin, and D.J. Singh, Phys. Rev. B {\bf 71} 214410 (2005)
(2004)
\end{thebibliography}
\end{document}